\documentclass{article}
\usepackage{amsmath,amsbsy,amsfonts,amssymb}
\usepackage{subfigure}
\usepackage{graphicx}
\usepackage{hyperref}

\usepackage[left=1.5cm,right=1.5cm,top=2cm,bottom=2cm]{geometry}

\author{Irving Rond\'on and F.~ Soto-Eguibar,\\
 Instituto Nacional de Astrof\'isica \'Optica y Electr\'onica,\\ Puebla, C.P. 72840 M\'exico.
}

\title{Properties of the Poynting vector for invariant beams:\\
	 negative propagation in Weber beams}
\begin{document}

\maketitle
\begin{abstract}
Negative propagation is an uncommon response produced by the local sign change in Poynting vector components. We present a general Poynting vector  expression  for all invariant beams with cylindrical symmetry using scalar potentials in order to evaluate the possibility of negative propagation. We  analyze the plausibility  of   negative  propagation being independent  of mode mixing;  we study Weber  beams as a particular case. The study of this negative effect allow us to advance in the field of micro manipulation and understanding of optical forces. Applications of these beams are discussed.	
\end{abstract}

\section{Introduction}
Propagation invariant beams, also known as \textquotedblleft non-diffracting beams\textquotedblright,  propagate indefinitely without changing their transverse intensity distribution. These invariant beams retain some of their peculiar characteristics: the basic transverse shape of  the field is preserved along their propagation. Some interesting properties  are absence of diffraction, self-reconstruction, highly focused field distributions  and angular momentum transfer. Novel problems have been addressed  and their applications  have been recently reported  in  \cite{Hugo}. These optical fields are the well known plane waves, Bessel beams \cite{Durnin}, Mathieu beams \cite{Julio} and Weber beams \cite{Bandres}.  The  study of these fields cover different areas, such as quantum optics \cite{MBandres1,RJ,BM1,BM2,BM3}, nonlinear optics \cite{Wulle}, optical communications \cite{Willner}, mechanical transfer of orbital angular momentum to trapped particles in optical tweezers \cite{Karem1},  trapping forces \cite{Hugo1}, acoustic optics  \cite{Zhang}, scattering problems using partial waves series in the far field approximation \cite{Marston1,Marston2,Chafiq,Chafiq2}, and angular momentum  \cite{Brandao} , among others.\\
In order to obtain the maximum possible information from these fields, it is very important to have a deeper insight of their physical properties. One of the key electromagnetic quantities is Poynting vector. The Bessel beams case has been successfully studied, both experimentally \cite{Lin} and theoretically \cite{Mokhun,Igor}. A complete study has also been done for Airy beams, first reported in \cite{Sztul} where the Poynting vector's evolution and angular momentum are analyzed. The authors in \cite{Novitsky} have recently shown  that Bessel beams  possess  negative values in the longitudinal and azimuthal components of the Poynting  vector, depending on  the phase detuning between the complex amplitudes $c_{TE}$ and $c_{TM}$ of the transversal electric part (TE) and the transversal magnetic part (TM) of the beam.  It should be noted that the Poynting vector's negative sign was investigated earlier for interference of four linearly polarized plane waves in  \cite{BZ}, but the author concluded that this fact does not have a realistic physical interpretation. Another interesting case was reported for X-Waves, where the propagation direction of their negative Poynting vector could be locally changed using carefully chosen complex amplitudes \cite{Salem}. Nonetheless, the negative Poynting vector  behavior presented  in  the Bessel beams and  in the X-Waves, mentioned above, has opened a discussion related to tractor beams generation, and other interesting  applications, such as the forces that  can be locally oriented in a direction opposite of the propagation wave vector \cite{Sukhov}. Examples of this last type of forces are: optical pulling force \cite{Jack}, negative effects in metamaterials \cite{JTCosta}, the optical forces process induced by nonparaxial gradient-less beams \cite{Qiu1}, the origin of universal optical traction \cite{Qiu2}, the application of  negative longitudinal optical force  under the illumination of  Bessel beam of arbitrary order, and polarization \cite{Wang}. On the other hand, in \cite{Pablo}, the authors found a negative propagation effect in the Airy nonparaxial beams  for which a complex superposition of Transverse Electric (TE) and Transverse Magnetic (TM) modes is not mandatory. Otherwise, for Bessel beams and for X-Waves beams, the local negativeness depends on the mixed modes TE/TM superposition.\\
Despite the considerable  research on optical invariant beams,  a general  description of  the Poynting vector in nonparaxial beams has not been presented, and the applications previously  mentioned  still remain unexplored for the Weber beam case. A need to understand their physical  properties  is necessary due to recent interesting works that have been reported using invariant beams, such as optical surfaces waves \cite{Tan,Marco}, application in laser beam shaping \cite{Arnold}, meta-surfaces \cite{Grbic}, optical forces \cite{Mitri}, scattering \cite{Hamed},  among others. Many of these works, however, are based on plane waves and attempts with Bessel beams. To the best of our knowledge,  very little attention has been reported for the Weber beams Poynting vector, considering any invariant beam and using  the scalar potential description. We propose an alternative  approach to achieve a description of the local negative properties for the Poynting vector. We show that mixed electromagnetic modes TE/TM are not a compulsory  condition to display  negative propagation for invariant beams. The transverse structure of the Weber beams can also be applied in \textquotedblleft tractor beams\textquotedblright  \cite{Dogariu}. Since the experiments carried out for the \textquotedblleft Bessel-tractor\textquotedblright case only works in short  distances \cite{Zemanek}, other applications such as optical and scattering forces can be explored \cite{Mazilu}.\\
This article is structured  as follows: in Section 2,  we  develop the theoretical framework based on the scalar potential approach; in Section 3,  we  present  the general  Poynting vector for all family invariant beams  and we give some examples,  as the plane wave and Bessel beam; in Section 4, we  consider the case of  Weber beams; and finally, in Section 5, we present our conclusions.

\section{The scalar potential as a general solution for Maxwell equations }\label{SectionScalarPotential}
It is well known  that  Maxwell's equations  can be written using a general vector basis  in a boundary value problem dealing with linear, isotropic, homogeneous, and time-invariant media, where these vector  eigenfunctions are derived from the complete set of separable solutions of the scalar Helmholtz  equation \cite{Boyer}. The method of coupling a pair of scalar functions with adequate boundary conditions for the electric and magnetic fields, was originally proposed to find solutions to the problem of  scattering of electromagnetic waves from a dielectric sphere. In \cite{Halas}, and references therein, the reader can find an interesting review of this topic and practical examples are given in \cite{Gumerov}.\\
In this work, we follow the formalism proposed by Stratton in \cite{Stratton} to study the properties of vector beams in cylindrical symmetry electromagnetic fields. In this formalism, the electromagnetic fields are written as
\begin{subequations}
	\begin{align}
	\vec{E}&=c_{TE}\vec{M}(\vec{r}) +  c_{TM} \vec{N}(\vec{r}), \label{ec:VecE} \\
	\vec{H}&=-i \sqrt{\frac{\varepsilon}{\mu}}\left[ c_{TE}\vec{N}(\vec{r}) + c_{TM} \vec{M}(\vec{r})\right], \label{ec:VecH}
	\end{align}
\end{subequations}
being $\vec{M}(\vec{r})$ and $\vec{N}(\vec{r})$ vector fields defined by
\begin{equation}
\label{ec:VecM}
\vec{M}(\vec{r})= \nabla \times [\hat{a} \psi(\vec{r})],
\end{equation}
and
\begin{equation}
\label{ec:VecN}
\vec{N}(\vec{r})= \frac{1}{k} \nabla \times \vec{M}(\vec{r}),
\end{equation}
where $\psi$ is a scalar field, $\hat{a}$ is an arbitrary unit vector  that determines the direction of propagation (which we will choose as the $Z$ axes, so $\hat{a}=\hat{e}_3$),  $k$ is the magnitude of the wave vector, $\varepsilon$ is the electric permittivity, $\mu$ is the magnetic permeability, and $c_{TE}$ and $c_{TM}$ are two arbitrary complex numbers \cite{Karem2}.\\
It is straightforward to verify that if the scalar field $\psi(\vec{r})$ satisfies the scalar Helmholtz equation,
\begin{equation}
\nabla^{2} \psi + k^{2} \psi =0,
\end{equation} 
then the fields (\ref{ec:VecE}) and (\ref{ec:VecH}) satisfy the Maxwell equations; so, the scalar field $\psi(\vec{r})$ will be  named scalar potential. Note that the vector fields, $\vec{M}$ and $\vec{N}$, are orthogonal between them, that is $\vec{M} \cdot \vec{N}=0$, and solenoidal, i.e. $\nabla \cdot \vec{M}=0$ and $\nabla \cdot \vec{N}=0 $.\\
Though the homogeneous (source-free) Helmholtz equation can be separated in eleven coordinate systems, we require separability into transverse and longitudinal parts, and that is possible only in Cartesian, cylindrical, parabolic cylindrical and elliptical cylindrical coordinates \cite{Boyer}. The spatial evolution of  the scalar potential $\psi$  can then be described by the transverse and the longitudinal parts. The transverse part $\varphi(u_1, u_2)$ will depend only on the transverse coordinates, $u_1,u_2$, and the longitudinal part $Z(z)$ will depend on the longitudinal coordinate $z$ (as we choose $\hat{a}=\hat{e}_3$). Therefore
\begin{equation}
\label{ec:VarSepa}
\psi(u,v,z)=\varphi(u_1,u_2)Z(z).
\end{equation}
After  substituting (\ref{ec:VarSepa}) in the Helmholtz equation, we easily obtain that $\varphi(u_1, u_2)$ satisfies the two dimensional transverse Helmholtz equation
\begin{equation} \label{ec:HemholtzTrans}
\nabla _T^2\varphi + k^2_T \varphi=0,
\end{equation}
where $\nabla _T^2$ is the Laplacian transversal operator which has a specific form in each coordinate system, and the longitudinal part is $Z(z)=e^{i k_z z}$ with the dispersion relation $k^2= k_T^2 + k_z^2$. As previously stated, the two dimensional transverse Helmholtz equation \eqref{ec:HemholtzTrans} can be separated in Cartesian, cylindrical, parabolic cylindrical and elliptical cylindrical coordinates \cite{Boyer}, and that gives origin to the plane waves, Bessel beams, Weber beams and Mathieu beams, respectively.  Then we can write
\begin{equation}
\label{ec:OpM}
\vec{M}=-e^{i k_z z}\nabla _T^{\bot }\varphi,
\end{equation}
where
\begin{equation}\label{nablatp}
\nabla _T^{\bot }=-\hat{e}_1\frac{1}{h_2}\frac{\partial }{\partial u_2}+\hat{e}_2\frac{1}{h_1}\frac{\partial }{\partial u_1},
\end{equation}
$ \hat{e}_1 $ and $\hat{e}_2$ are the base unit vectors corresponding to the transversal direction, and $h_1$ and $h_2$ are the corresponding scale factors.  We note that in the four coordinate systems we are studying, the scale factor $ h_3$ is equal to 1. It is also very easy to see that
\begin{equation}
\label{ec:OpN}
\vec{N}= \frac{e^{i k_z z}}{k}\left(  i k_z \nabla _T+\hat{e}_3 k_T^2\right)\varphi,
\end{equation}
where
\begin{equation}\label{nablat}
\nabla _T=\hat{e}_1\frac{1}{h_1}\frac{\partial }{\partial u_1}+\hat{e}_2\frac{1}{h_2}\frac{\partial }{\partial u_2}.
\end{equation}
It is important to note that the scalar potential can be used to analyze any  invariant beam. In the case of Bessel beams, this formalism was thoroughly used for analyzing them theoretically in \cite{Mishra, Olivik,Bouchal}, and in \cite{Flores} to study experimentally its TE and TM modes.\\
In the particular case of Weber beams, its transversal structure is  naturally described in terms of parabolic coordinates $ u$ and $v$,  related to the Cartesian through the following transformation
\begin{align}
\label{ecCarTranf}
x + iy &= \frac{1}{2}(u+iv)^2 , \qquad  z=z ,
\end{align}
where   $x$, $y$ and $z$ are the corresponding Cartesian coordinates with $ u \in (-\infty, + \infty) $ and $ v \in [0, +\infty) $. The scaling factors for the parabolic coordinate system are  given by 
\begin{align}
h_{1}=h_{2}=  \sqrt{u^2 + v^2}.
\end{align}

\section{ A general expression  for the transversal and  longitudinal beam propagation modes}
The Poynting vector represents the directional power flux per unit area of an electromagnetic field. For harmonic electromagnetic fields, its time average is given by 
\cite{Jackson}  
\begin{equation} \label{ec:VecPoyn}
\left\langle  \vec{S}  \right\rangle= \frac{1}{2} \mathrm{Re} \left(\vec{E} \times \vec{H}^* \right).
\end{equation}
Writing the electromagnetic fields, \eqref{ec:VecE} and \eqref{ec:VecH}, in terms of the scalar potential $\varphi$, we can obtain the time averaged Poynting vector for any invariant beams as a sum of a transversal and a longitudinal part,
\begin{equation}
\label{ec:STSZ}
\left\langle  \vec{S} \right\rangle =  \left\langle  \vec{S} \right\rangle_{\text{T}} +  \left\langle  \vec{S} \right\rangle _{\text{z}},
\end{equation}
where the transversal Poynting vector is
\begin{align}
\label{ec:St} 
\left\langle  \vec{S} \right\rangle_{\text{T}} &= \frac{k_T^2}{2k^2} \sqrt{\frac{\varepsilon}{\mu}} \times
\nonumber \\
&\mathrm{Re} \left[  
-i  \left(   \left| c_{\text{TE}}\right| ^2    \varphi^\ast\nabla_T \varphi -   \left|  c_{\text{TM}}  \right|^2    \varphi\nabla_T \varphi^\ast    \right)     k
+  k_z  c_{\text{TE}}^\ast  c_{\text{TM}}  \hat{e}_3 \times \nabla_T   \left(  \varphi^* \varphi   \right) 
\right] ,
\end{align}
and the longitudinal Poynting vector is 
\begin{align}
\label{ec:Sz}
\left\langle  \vec{S}   \right\rangle_{\text{z}} &= \frac{\hat{e}_3}{2k^2} \sqrt{\frac{\varepsilon}{\mu}}  \,  
\mathrm{Re}
\Bigg[
\left(   \left|c_{\text{TE}}\right|^2 +\left|  c_{\text{TM}}\right|^2 \right)  k k_z
\left(   \nabla_T  \varphi \cdot \nabla_T \varphi^\ast    \right)
\nonumber \\ &
+i\left( c_{\text{TE}}c_{\text{TM}}^\ast k^2+c_{\text{TE}}^\ast c_{\text{TM}} k_z^2  \right) 
\left(     \nabla_T \varphi^*\times \nabla_T \varphi  \right)    \cdot  \hat{e}_3 \Bigg]
\end{align} 
The expressions  \eqref{ec:St} and \eqref{ec:Sz}  give all the information about the Poynting vector for any invariant beam, and they provide  information related with a particular mode TE or TM and the possible mixed modes. These expressions posses an interference part which generally is not zero. Note that these results show  that the time averaged Poynting vector is independent of the $z$ coordinate, as expected. In order to emphasize the non-diffractive character of these beams, it  was proven  that the divergence  of the transversal part of the time averaged Poynting vector is zero \cite{Horak}. All the previous argumentations show that when mixed modes are present, the acquisition of electromagnetic properties is straightforward using the scalar approach \cite{IRondon}.\\
The next step is to generalize the procedure proposed in \cite{Novitsky}, for Bessel beams, to any invariant beam. The substitution of $c_{\text{TE}}= \vert  a_1 \vert  e^{i \phi_1}$ and $c_{\text{TM}}= \vert a_2 \vert  e^{i \phi_2}$ in  the equations  \eqref{ec:St} and \eqref{ec:Sz} shows directly that any invariant beam has negative behavior due to mixed modes; mathematically, a periodic function cosine or sine, with certain phase differences,  can be presented,  as was reported for Bessel beams \cite{Novitsky}. However, recently it has been stated in  \cite{Pablo} that the negative behavior can be found independently of the mode interference. In order to validate the possibility of negative local Poynting vector behavior,  let us  use equations \eqref{ec:St} and \eqref{ec:Sz} in the mixed  components, in the following examples. \\
In the case of a plane wave, where  the solution of the transversal Helmholtz equation is $ \varphi(x,y) = \exp{[i(k_x x + k_y y )]}$, is very easy to show that  $  \mathrm{Re}[c^{*}_{\text{TE}} c_{\text{TM}}\hat{e}_3 \times \nabla_T (\varphi^* \varphi) ]=0 $,  and that $\mathrm{Re}[ i\left(c_{\text{TE}}c_{\text{TM}}^\ast k^2+c_{\text{TE}}^\ast c_{\text{TM}} k_z^2\right)\left(  \nabla_T \varphi^*\times \nabla_T \varphi \right)] =0$; therefore, the interference term  is zero. However, negative propagation can be found using several plane waves; indeed, two plane wave are sufficient \cite{BZ}, without  mixed modes. \\
For the Bessel beams, it is very well known that the transversal field is $ \varphi(r, \theta) = J_m (k_T r)\exp({i m \theta })$; and in this case we have that $  \mathrm{Re}[c^{*}_{\text{TE}} c_{\text{TM}}\hat{e}_3 \times \nabla_T (\varphi^* \varphi) ]$ and $\mathrm{Re}\left[  i\left(c_{\text{TE}}c_{\text{TM}}^\ast k^2+c_{\text{TE}}^\ast c_{\text{TM}} k_z^2\right)\left(  \nabla_T \varphi^*\times \nabla_T \varphi \right)\right] $ are both different from zero, which implies negative propagation in the azimuthal and longitudinal components \cite{Novitsky}. In the next section, the hitherto unreported negative behavior of the Poynting vector in the case of Weber beams will be considered. 

\section{The Poynting vector of Weber beams}
\begin{figure}
	\centering
 \includegraphics[scale=0.6]{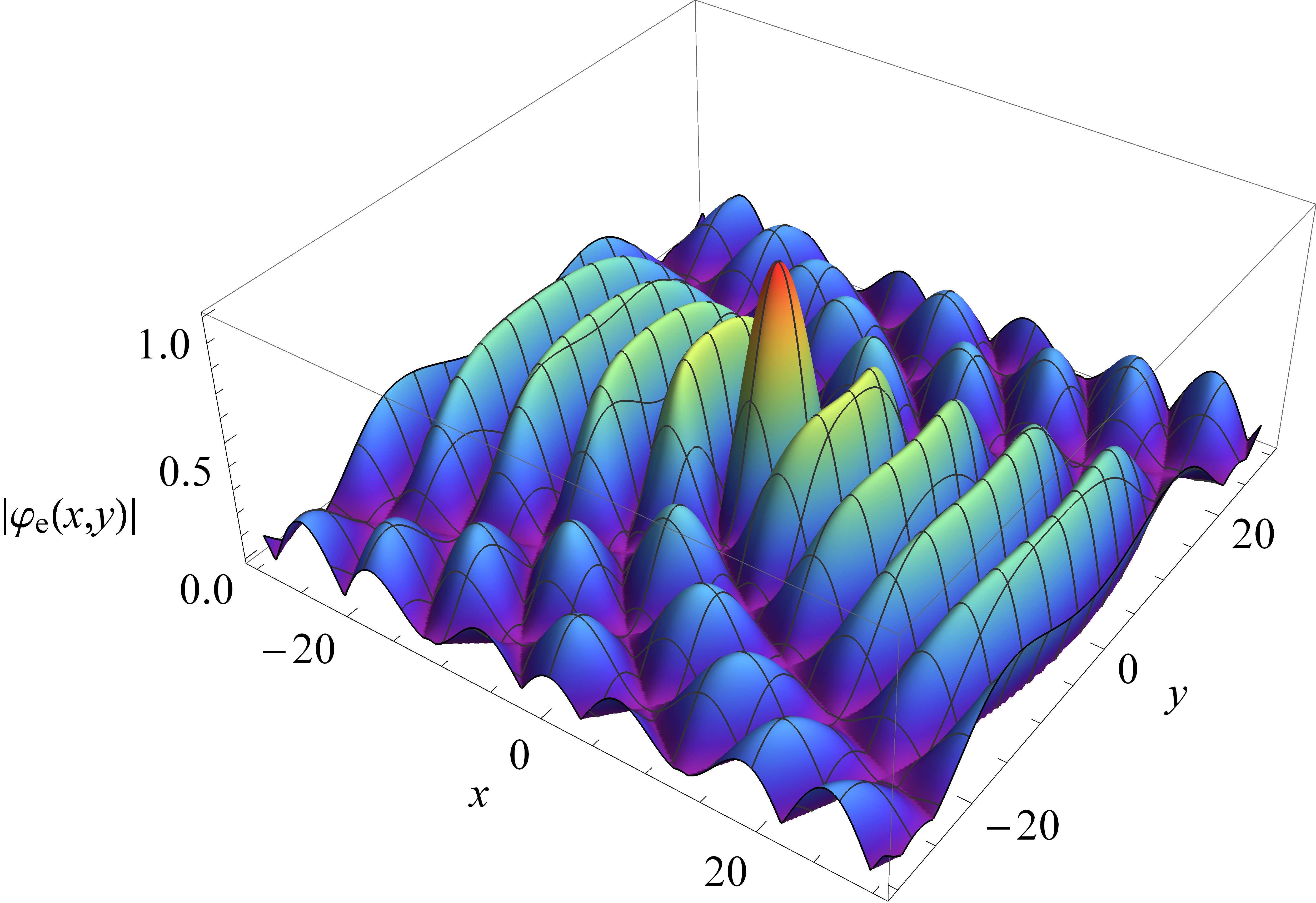}
\includegraphics[scale=0.6]{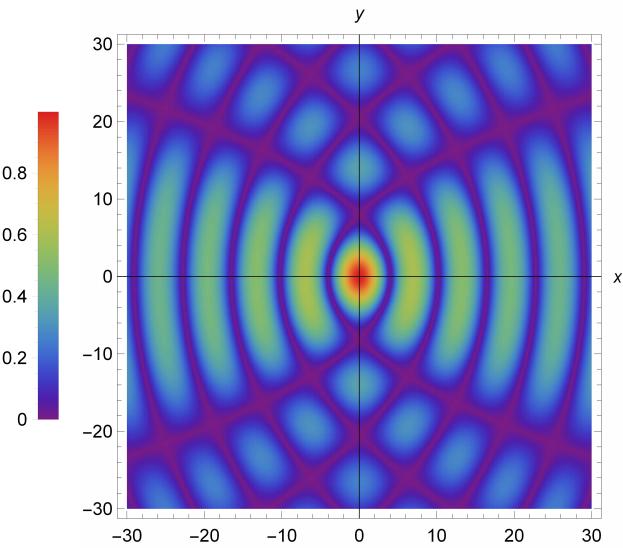}
	\caption{Absolute value of the even Weber beams \eqref{ec:WeberPar} for $a=0$ and $k_T= 1\; \mathrm{m}^{-1}$. Colors represent the absolute value of the even Weber beams according the bar shown.}
	\label{Figure1}
\end{figure}

\begin{figure}
	\centering
\includegraphics[scale=0.6]{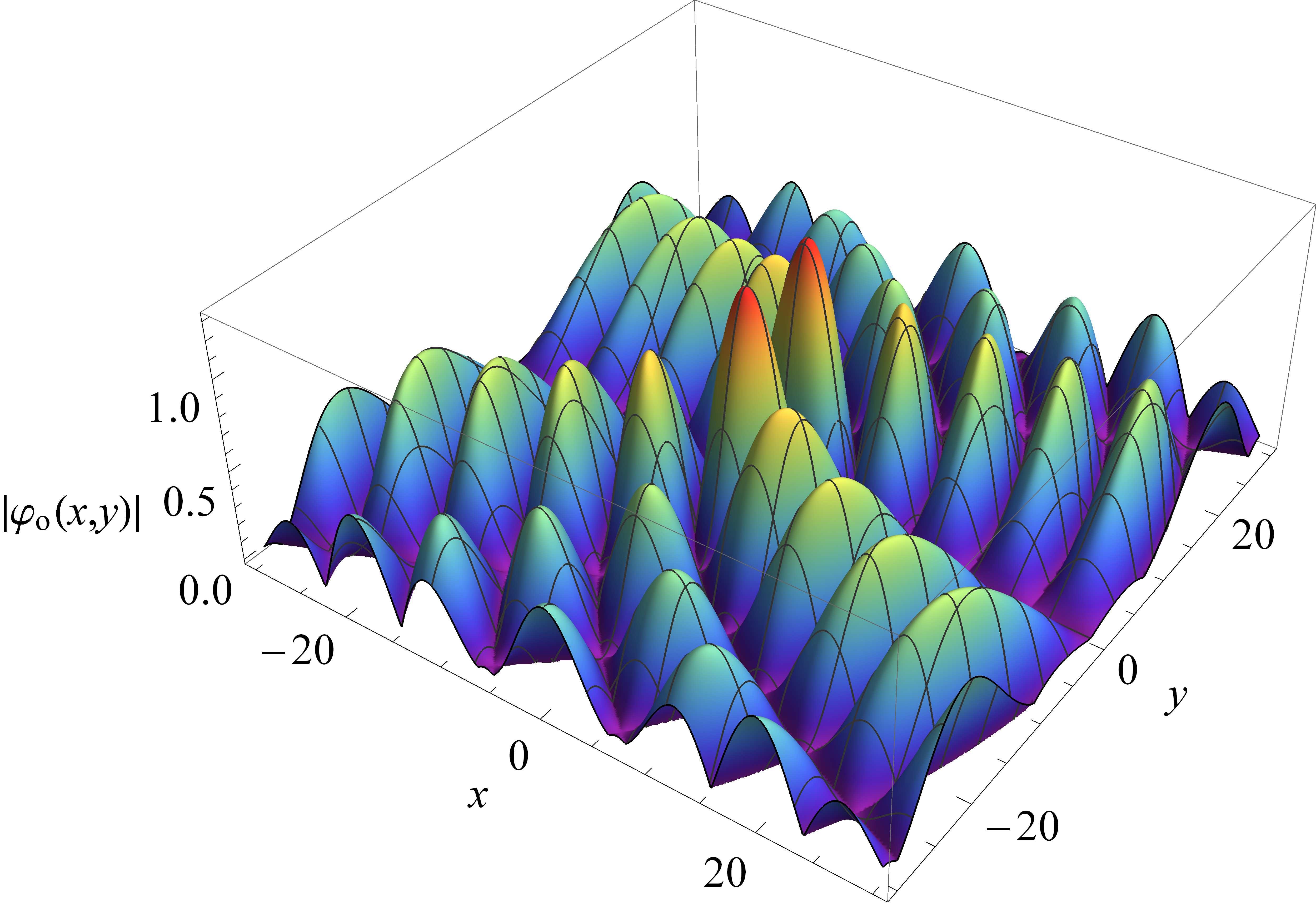}
\includegraphics[scale=0.6]{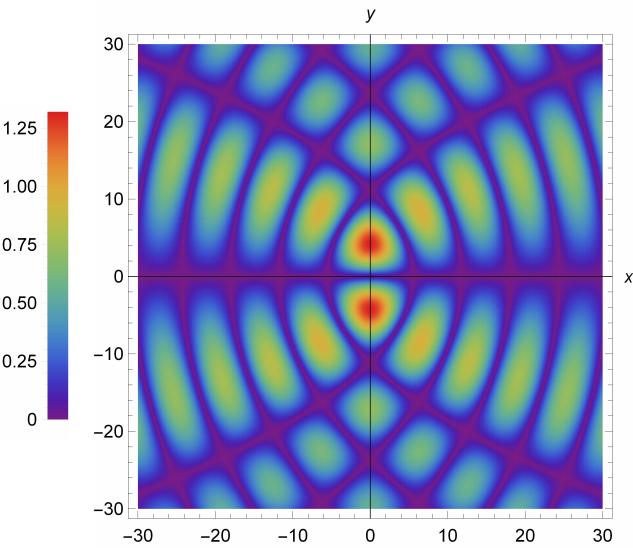}
	\caption{Absolute value of the odd Weber beams \eqref{ec:WeberImpar} for $a=0$ and $k_T= 1\; \mathrm{m}^{-1}$. The colors represent the absolute value of the odd Weber beams according the bar shown.}
	\label{Figure2}
\end{figure}

\begin{figure}  
	\centering
\includegraphics[scale=0.8]{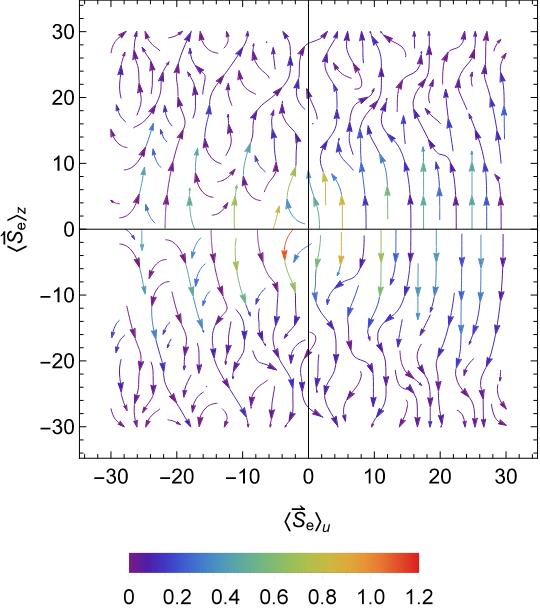}
\includegraphics[scale=0.8]{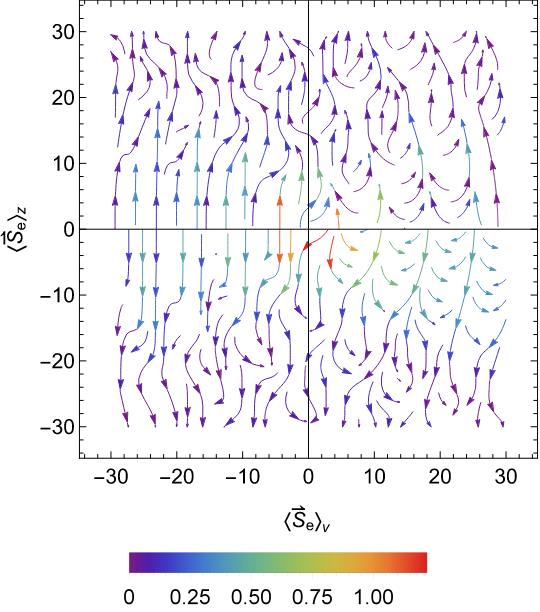}
\caption{Streamlines of the two dimensional vector fields
$ \left( \left\langle  \vec{S}_{\text{e}} \right\rangle_u,\left\langle  \vec{S}_{\text{e}} \right\rangle_z \right)  $
and
$\left( \left\langle  \vec{S}_{\text{e}} \right\rangle_v,\left\langle  \vec{S}_{\text{e}} \right\rangle_z \right)  $,
for $a=0$, $k_T= 1\;  \mathrm{m}^{-1}$ and $\phi_1 = \phi_2$. The plots streamlines show the local direction of the vector field at each point and the arrows are colored according to the magnitude of the field, as presented in the bars below. }
\label{Fig:ProyVecPoynWpar}	
\end{figure}

\begin{figure}    
	\centering
\includegraphics[scale=0.8]{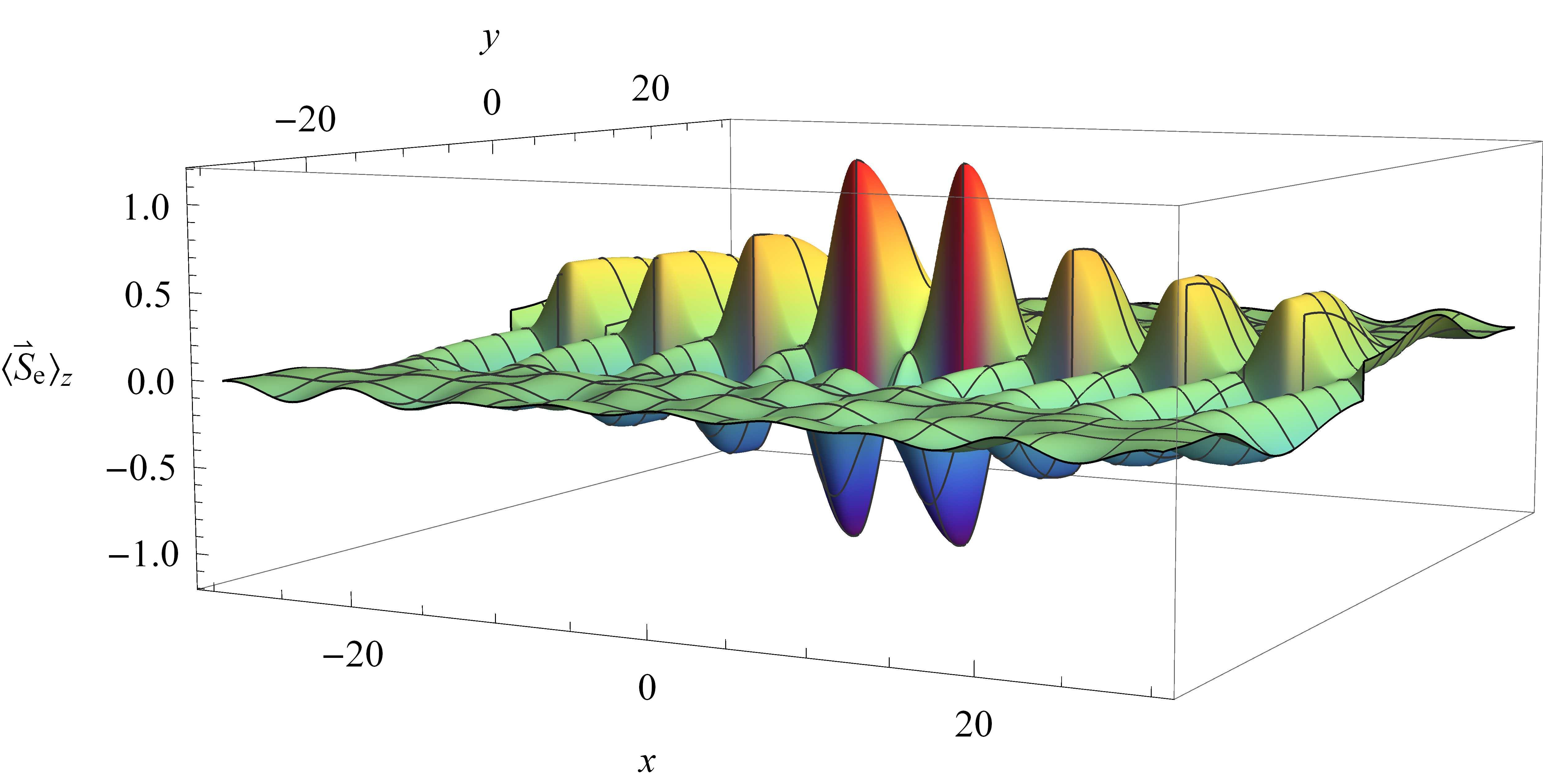}
\includegraphics[scale=0.6]{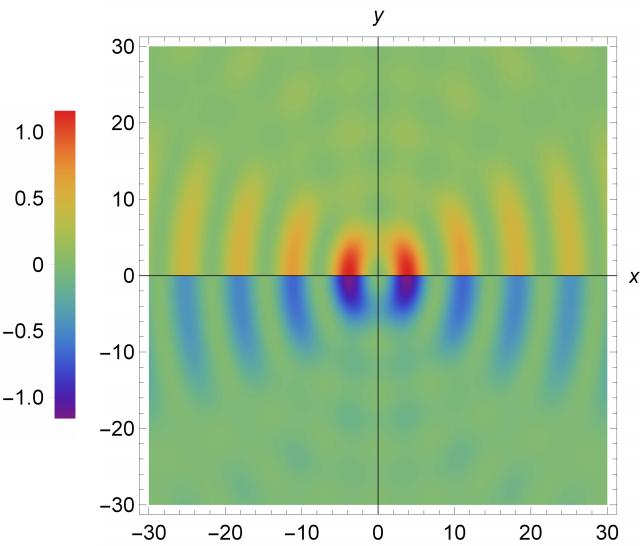}
	\caption{Tridimensional behavior of the longitudinal component of the Poynting vector, $\left\langle  \vec{S}_{\text{e}} \right\rangle_z$, for even Weber beams ($a=0$, $k_T= 1 \; \mathrm{m}^{-1}$ ). The colors represent the values of the $z-$component of the Poynting vector, according to the bar shown. }
	\label{Fig:3D2DSzWpar}
\end{figure}

\begin{figure} 
\centering
\includegraphics[scale=0.8]{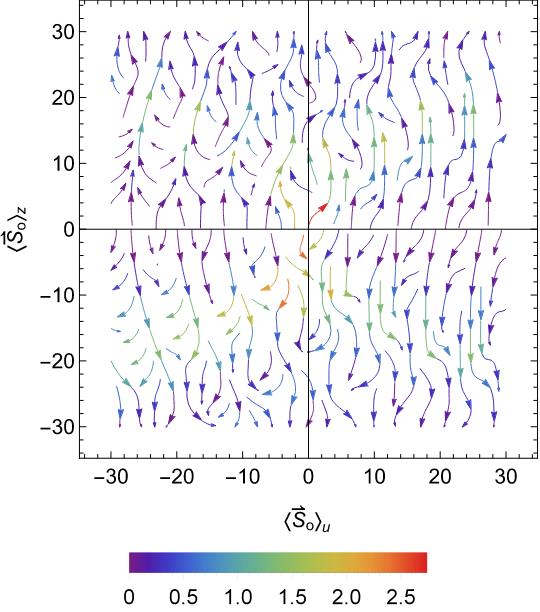}\quad
\includegraphics[scale=0.8]{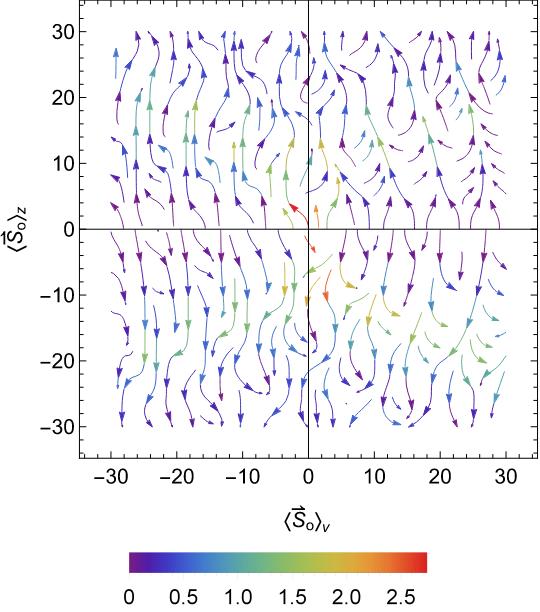}
\caption{ Streamlines of the two dimensional vector fields
		$ \left( \left\langle  \vec{S}_{\text{o}} \right\rangle_u,\left\langle  \vec{S}_{\text{o}} \right\rangle_z \right)  $
		and
		$\left( \left\langle  \vec{S}_{\text{o}} \right\rangle_v,\left\langle  \vec{S}_{\text{o}} \right\rangle_z \right)  $,
		for $a=0$, $k_T= 1\;  \mathrm{m}^{-1}$ and $\phi_1 = \phi_2$. The plots streamlines show the local direction of the vector field at each point and the arrows are colored according to the magnitude of the field, as presented in the bars below. }
	\label{Fig:ProyVecPoynWimpar}
\end{figure}

\begin{figure}   
	\centering
\includegraphics[scale=0.8]{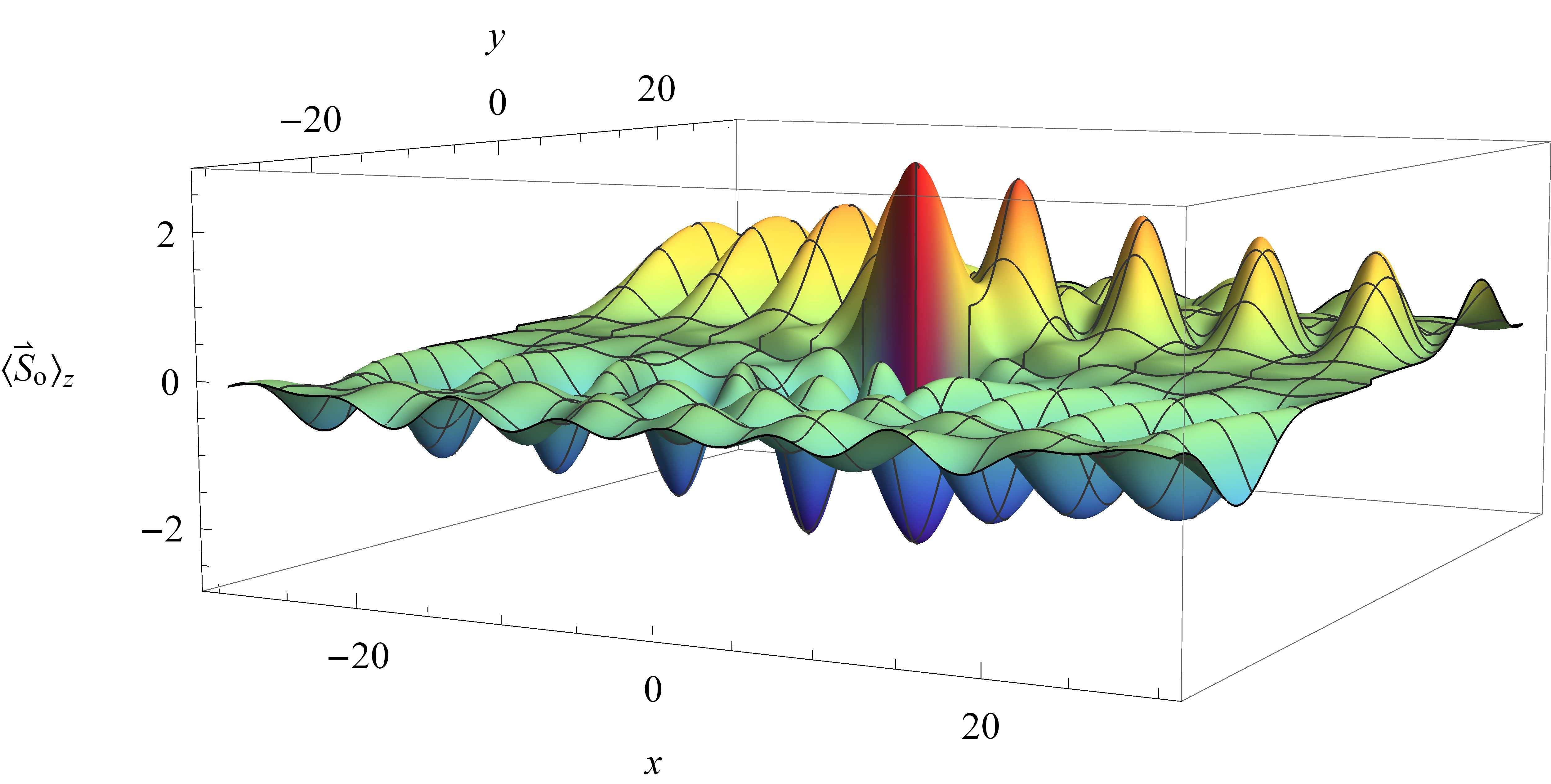}
\includegraphics[scale=0.6]{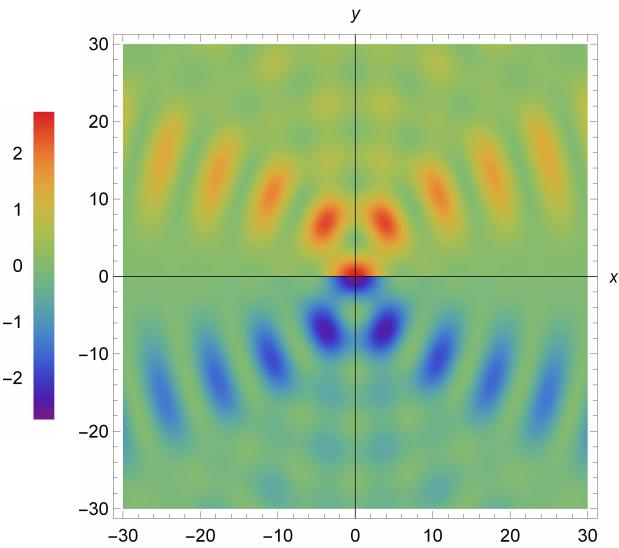}
	\caption{Tridimensional behavior of the longitudinal component of the Poynting vector, $\left\langle  \vec{S}_{\text{o}} \right\rangle_z$, for odd Weber beams ($a=0$, $k_T= 1 \; \mathrm{m}^{-1}$ ). The colors represent the values of the $z-$component of the Poynting vector, according to the bar shown. }  \label{Fig:3D2DSzWimpar}	
\end{figure}

The existence of parabolic optical fields has been  demonstrated theoretically and experimentally  for the first time by Bandres et. al. \cite{Bandres, MBandres1}. At the present time, several interesting investigations have been done, such as quantum experimental observation of the scattering of free falling dilute thermal clouds of $^{38}$Rb atoms \cite{JRocio} and the manipulation of microscopic particles driven by parabolic beams \cite{Petrov}.  The physical properties of Weber beams have been recently studied, such as accelerating beams \cite{MBandres2, Jeffrey}. Their implications for many linear wave systems in nature and its applications in diverse scenarios have been evaluated \cite{MBandres3}.\\
The scalar potential of the Weber beams can be expressed as a product of parabolic functions; for the even Weber beams,
\begin{align}
\label{ec:WeberPar}
\varphi_{\mathrm{e}}(u,v)=\frac{1}{\pi \sqrt{2}}\left| \Gamma_1\right|^2 U_{\mathrm{e}}(u,a) U_{\mathrm{e}}(v,-a),
\end{align}
and for the odd Weber beams,
\begin{align}
\label{ec:WeberImpar}
\varphi_{\mathrm{o}}(u,v)=\frac{1}{\pi \sqrt{2}}\left| \Gamma_3\right|^2 U_{\mathrm{o}}(u,a) U_{\mathrm{o}}(v,-a),
\end{align}
where the subscripts  e and o stand  for even and odd, respectively; $a$ is essentially the separation constant, that we will consider real,  $\Gamma_1=\Gamma(\frac{1}{4}+i \frac{a}{2})$ and $\Gamma_3=\Gamma(\frac{3}{4}+i \frac{a}{2})$  are normalization constants, being $\Gamma\left( \zeta \right)$ the gamma function,  and the $U$ functions are given by
\begin{equation}
\label{ec:UpWeber}
U_{\mathrm{e}}(u,a)= \exp  \left(-i \frac{1}{2} k_T  u^2 \right) {}_1 F_{1}\left(\frac{1}{4}-i \frac{a}{2},\frac{1}{2},i  k_T    u^2\right),
\end{equation}
and by 
\begin{equation}
\label{ec:UimpWeber}
U_{\mathrm{o}}(u,a)= u \exp  \left(-i \frac{1}{2} k_T  u^2 \right) {}_1 F_{1}\left(\frac{3}{4}-i \frac{a}{2},\frac{3}{2},i  k_T    u^2\right),
\end{equation}
with $ {}_1 F_{1}\left(\alpha,\beta, \zeta \right)  $ the confluent hypergeometric function. In Figures \ref{Figure1}  and \ref{Figure2}, the absolute value of the Weber beams, even \eqref{ec:WeberPar} and odd \eqref{ec:WeberImpar}, are shown for $a=0$ and $k_T= 1\; \mathrm{m}^{-1}$, in terms of the Cartesian coordinates.
\noindent
If we substitute equations \eqref{ec:WeberPar} to \eqref{ec:UimpWeber} in \eqref{ec:St} and  \eqref{ec:Sz}, we obtain, after a very long calculation, the general Poynting vector for the even and odd Weber beams; in this Poynting vector the transversal and longitudinal interference terms are not zero, thus there will be negative propagation. To simplify the analysis, we will consider here only the case $a = 0$; we have chosen this value for the sake of simplicity and because these particular Weber beams have been used to study the scattering by a rigid sphere centered along the propagation axis   \cite{Chafiq2}, to analyze the transversal behavior and for inspect some features of their normalization \cite{Blas3}. Thus, when $a=0$, we obtain for the even Weber beams,
\begin{subequations}
	\begin{align}
	\label{ec:Sxpar}
	\left\langle  \vec{S}_{\text{e}} \right\rangle_u = & \frac{  k_T^4  }{4 k^2} \sqrt{\frac{\varepsilon }{\mu}} \frac{k_z  v^2    }{ \sqrt{u^2+v^2}}  
	\Gamma ^4 \left(\frac{3}{4}\right) \left| u\right|
	J^2_{-\frac{1}{4}}\left(\frac{k_T u^2}{2}\right) J_{-\frac{1}{4}}\left(\frac{k_T v^2}{2}\right) 	\times
	\nonumber \\ &
	J_{\frac{3}{4}}\left(\frac{k_T v^2}{2}\right) \cos (\phi_1 - \phi_2),\\
	\label{ec:Sypar}
	\left\langle  \vec{S}_{\text{e}} \right\rangle_v &=
	-   \frac{ k_T^4 }{4 k^2} \sqrt{\frac{\varepsilon }{\mu}}
	\frac{  k_z  u v    }{ \sqrt{u^2+v^2}}  \Gamma ^4 \left(\frac{3}{4}\right) \left| u\right|  J_{-\frac{1}{4}}\left(\frac{k_T u^2}{2}\right) J_{\frac{3}{4}}\left(\frac{k_T u^2}{2}\right) \times
	\nonumber \\ &
	J^2_{-\frac{1}{4}}\left(\frac{k_T v^2}{2}\right) \cos (\phi_1 - \phi_2),\\
	\label{ec:Szpar}
	\left\langle  \vec{S}_{\text{e}} \right\rangle_z & =
	\frac{k_T^3 k_z   }{4 k}    \sqrt{\frac{\varepsilon }{\mu }}   \frac{v  \left| u\right|  }{ u^2+v^2}     \Gamma^4 \left(\frac{3}{4}\right) \times
	\nonumber \\ &
	\left[ v^2 J^2_{-\frac{1}{4}}\left(\frac{k_T u^2}{2}\right) J^2_{\frac{3}{4}}\left(\frac{k_T v^2}{2}\right)+u^2 J^2_{\frac{3}{4}}\left(\frac{ k_T u^2}{2}\right) J^2_{-\frac{1}{4}}\left(\frac{k_Tv^2}{2}\right)\right] .
	\end{align}
\end{subequations}
We can observe  the presence of spatial variations for $\left\langle  \vec{S}_{\text{e}} \right\rangle_u$ and $\left\langle  \vec{S}_{\text{e}} \right\rangle_v$ due to the phase difference between modes, as was reported for the Bessel beam case in \cite{Novitsky}. However, even when the phase difference is zero, $\phi_1=\phi_2$, we have negative values for all three components of the Poynting vector. 
The Poynting vector of the Weber beams has a very complex vector structure. In order to give an idea of this complex comportment, we show in Figure \eqref{Fig:ProyVecPoynWpar} the streamlines of the two dimensional vector fields
$ \left( \left\langle  \vec{S}_{\text{e}} \right\rangle_u,\left\langle  \vec{S}_{\text{e}} \right\rangle_z \right)  $
and
$\left( \left\langle  \vec{S}_{\text{e}} \right\rangle_v,\left\langle  \vec{S}_{\text{e}} \right\rangle_z \right)  $,
obtained as combinations of the two transverse components with the longitudinal component; the plots streamlines show the local direction of the vector field at each point and the arrows are colored according to the magnitude of the field, as presented in the bars below each one. The behavior of the trajectories and their relation with the parameters can be explained from the point of view of  dynamical systems \cite{Barkovsky}, but that goes beyond the scope of this work. 
\noindent As we already mention, these results show that for the longitudinal  component vector, given by \eqref{ec:Szpar}, there is not presence of mixed modes, which was the hypothesis for negative propagation in \cite{Novitsky}. The negative propagation for the longitudinal component is shown in Figure  \eqref{Fig:3D2DSzWpar}.
\noindent Following the same procedure, we found for the odd case,
\begin{subequations}
	\begin{align}
	\left\langle  \vec{S}_{\text{o}} \right\rangle_u  = &
	- \frac{ k_T^2 k_z \left| v\right|}{ k^2\left| u\right|}\sqrt{\frac{\varepsilon }{\mu}} \frac{u^2 v}{ \sqrt{ u^2+v^2}}  \Gamma \left(\frac{1}{4}\right) \Gamma \left(\frac{5}{4}\right)^3 J_{\frac{1}{4}}^2\left(\frac{k_T u^2}{2}\right) \times
	\nonumber \\ &
	\label{ec:Sximpar}
	J_{-\frac{3}{4}}\left(\frac{ k_T v^2}{2}\right) J_{\frac{1}{4}}\left(\frac{k_T v^2}{2}\right) \cos  (\phi_1-\phi_2)\\
	\left\langle  \vec{S}_{\text{o}} \right\rangle_v  & =
	\frac{ k_T^2 k_z  }{k^2 \left| u\right| } 
	\sqrt{\frac{\varepsilon }{\mu }} 
	\frac{ u^3 v  }{\sqrt{u^2+v^2}} 
	\Gamma \left(\frac{1}{4}\right) \Gamma \left(\frac{5}{4}\right)^3  
	J_{-\frac{3}{4}}\left(\frac{k_T u^2}{2}\right) \times
	\nonumber \\ &
	\label{ec:Syimpar}
	J_{\frac{1}{4}}\left(\frac{k_T u^2}{2}\right) J^2_{\frac{1}{4}}\left(\frac{k_Tv^2}{2}\right)
	\cos (\phi_1-\phi_2) , \\
	\left\langle  \vec{S}_{\text{o}} \right\rangle_z  & =
	\frac{4 k_T k_z     }{k}     \sqrt{\frac{\varepsilon }{\mu }} \frac{v  \left| u\right|   }{u^2+v^2}   \Gamma^4 \left(\frac{5}{4}\right) \times
	\nonumber \\ &
	\left[ u^2 J^2_{-\frac{3}{4}}\left(\frac{k_T u^2}{2}\right) J^2_{\frac{1}{4}}\left(\frac{k_T v^2}{2}\right)+v^2 J^2_{\frac{1}{4}}\left(\frac{k_T u^2}{2}\right) J^2_{-\frac{3}{4}}\left(\frac{k_T v^2}{2}\right)\right] .
	\end{align}	
\end{subequations}
As in the even case, the negative propagation appears even if there are no mixed modes.
As expected, the behavior of the Poynting vector in this case is as complex as in the even case, and to visualize it we present Figure  \eqref{Fig:ProyVecPoynWimpar}
where the streamlines of the two dimensional vector fields
$ \left( \left\langle  \vec{S}_{\text{o}} \right\rangle_u,\left\langle  \vec{S}_{\text{o}} \right\rangle_z \right)  $
and
$\left( \left\langle  \vec{S}_{\text{o}} \right\rangle_v,\left\langle  \vec{S}_{\text{o}} \right\rangle_z \right)  $,
obtained as combinations of the two transverse components with the longitudinal component, are plotted; the plots streamlines show the local direction of the vector field at each point and the arrows are colored according to the magnitude of the field, as presented in the bars below each one. The negative energy propagation in the longitudinal component  $\left\langle  \vec{S}_{\text{o}}  \right\rangle_z$ of the odd Weber beams is shown in Figure \eqref{Fig:3D2DSzWimpar}.

\section{Conclusions}
We have obtained  a general expression for the Poynting vector using a scalar approach, and we have shown  its  potential  to study the energy negative behavior for any invariant beam with cylindrical symmetry. The  expression obtained for the Poynting vector brings to the fact that all nonparaxial  beams have the possibility of negative local change in the Poynting vector, independently of the presence of mixed modes. In the case of Weber beams, we have shown explicitly   that negative propagation exists. In the longitudinal direction, $S_z <0$ independently of the difference between  $c_{\text{TE}}$ and $c_{\text{TM}}$ for $a=0$.  The presence of negative propagation regions in Weber beams  can  provide new applications, such as trapping, because  the variation of the Poynting vector sign can create multiple traps and bounding particles. Recently the authors in \cite{Cheng} have reported the use of optical engineering and  light-matter interaction for emerging optical manipulation and its applications, but the  Weber beam case still needs to be explored.

\section{Acknowledgments}
\noindent
I. Rondon-Ojeda  thanks and acknowledges the support given by SEP-PRODEP postdoctoral program UMSNH-CA-221 at Laboratory of Optical Sensors. The authors also want to thank Douglas David Crockett, Volunteer of Peace Corps Response at INAOE, for reading and improve the manuscript . We also want to thank the unknown reviewers and the editor of this journal for several valuable corrections and comments.

\end{document}